\documentclass{article}
\usepackage{amssymb}
\usepackage{amsmath}
\usepackage{graphicx}
\usepackage{amscd,multicol}

\title{Extra generations and discrepancies of electroweak precision data}
\author{V.A. Novikov, L.B. Okun, \\
ITEP, Moscow, Russia \\
 A.N. Rozanov, \\
CPPM, IN2P3, CNRS, Univ. Mediteranee, Marseilles, France \\
 and
ITEP, Moscow, Russia
\\
  M.I. Vysotsky \\
ITEP, Moscow, Russia  }
\date{}
\begin{document}
\maketitle

\begin{abstract}

It is shown that additional chiral generations are not excluded by
the latest electroweak precision data if one assumes that there
is no mixing with the known three generations. In the case of ``heavy
extra generations'', when all four new particles are heavier than
$Z$ boson, quality of the fit for the one new generation is as
good as for zero new generations (Standard Model). In the case of
neutral leptons with masses around 50 GeV (``partially heavy
extra generations'') the  minimum of $\chi^2$ is
between one and two extra generations.
\end{abstract}

Two years ago in paper \cite{1} we analyzed bounds from the
electroweak precision data on the
non-decoupled New Physics in a form of additional heavy
quark-lepton generations.  It was shown that while the
case of all four new fermions ($U$ and $D$ quarks, neutral lepton
$N$ and charged lepton $E$) heavier than $Z$ boson was excluded at
2.5 $\sigma$ level, existence of new generations with relatively
light neutral lepton $N$ ($m_N \approx 50$ GeV) was allowed.
At that time quality of
Standard Model (SM) fit of the data was very good, $\chi^2/n_{\rm
d.o.f.} = 15/14$. At the time of Osaka Conference, summer 2000,
nothing radical happened but $\chi^2$ became $21/13$ and the level
at which one extra heavy generation was excluded went down to
$2\sigma$ \cite{2}. However the latest precision data announced
summer 2001  \cite{3} has changed the situation: the fit is still
bad, $24/13$, but now the presence of one heavy generation does
not make the fit worse as compared with SM. \maketitle
\begin{table}[t]
\caption {LEPTOP fit to electroweak observables. Year 2001. By
italics we designate calculated (not measured) quantities.
     }

\vspace{5mm}
\begin{center}

\begin{tabular}{|l|l|r|r|r|}
\hline
Source & Observable & Exp. data & LEPTOP fit & Pull \\ \hline
& $\Gamma_Z$ [GeV] &    2.4952(23) &  2.4966(16)  & -0.6   \\
& $\sigma_h$ [nb] &   41.540(37)& 41.480(14)  & 1.6    \\
& $A_{FB}^l$ & 0.0171(10)  & 0.0165(3)  & 0.7   \\
& $R_l$ & 20.767(25)& 20.738(18)  & 1.1   \\
LEP I & $A_\tau$, $A_e$ & 0.1465(33) & 0.1483(11) &  -0.5 \\
& $R_b$ &    0.2165(7) &   0.2157(1)  & 1.2 \\
& $R_c$    &    0.1719(31)&   0.1723(1)  & -0.1   \\
& $A_{FB}^b$ &    0.0990(17)&   0.1040(8)  & -2.9   \\
& $A_{FB}^c$  & 0.0685(34)&   0.0743(6)  & -1.7   \\
& $s_l^2(Q_{FB})$   & 0.2324(12)  &   0.2314(1)  & 0.9   \\
\hline
& $A_{LR}$   & 0.1513(21)  &   0.1483(11)  & 1.4   \\
SLC & $s_l^2$ ($A_{LR}$) & {\it 0.2310(3)} &   {\it 0.2314(1)}  & -1.4   \\
& $A_b$ &  0.9220(200) & 0.9349(1)  & -0.6   \\
& $A_c$ &  0.6700(260)&   0.6684(5)  & 0.1 \\
\hline
LEP II, Tevatron & $m_W$ [GeV] & 80.451(33) & 80.392(20)  & 1.8   \\
& $s_W^2(m_W)$ & {\it 0.2216(6)} & & \\
\hline
Tevatron & $s_W^2$ ($\nu N$) & 0.2255(21)& 0.2230(3) & 1.2 \\
& $m_W(\nu N)$ [GeV] & {\it 80.250(109)} & & \\
& $m_t$ [GeV]    & 174.3(5.1) &    175.0(4.4) & -0.1\\
\hline
Fit & $m_H$ [GeV]    & &  { $79^{+47}_{-29} $} &  \\
& $\hat{\alpha}_s$ &           & 0.1182(27) &  \\
\hline
$e^+ e^- \to$ hadrons & $\bar{\alpha}^{-1}$ & 128.936(49)   &
128.918(45) & 0.4  \\
\hline
& {\small $\chi^2/n_{d.o.f.}$} & & 23.8/13 &
\\ \hline
\end{tabular}
\end{center}
\end{table}

\begin{table}[t]
\caption {LEPTOP fit to electroweak observables. Year 2000.
     }

\begin{center}
\begin{tabular}{|l|l|r|r|r|}
\hline
Source & Observable & Exp. data & LEPTOP fit & Pull \\
\hline & $\Gamma_Z$ [GeV] &    2.4952(23) &  2.4964(16)  & -0.5
\\ & $\sigma_h$ [nb] &   41.541(37)& 41.479(15)  & 1.7    \\
& $A_{FB}^l$ & 0.0171(10)  &  0.0164(3)  & 0.7   \\
& $R_l$ &
20.767(25)& 20.739(18)  & 1.1   \\ LEP I & $A_\tau$, $A_e$ &
0.1467(32) &  0.1480(13)  & -0.4   \\ & $R_b$ &    0.2165(7) &
0.2157(1)  & 1.2   \\ & $R_c$    & 0.1709(34)&   0.1723(1)  & -0.4
\\ & $A_{FB}^b$  &    0.0990(20)& 0.1038(9)  & -2.4   \\ &
$A_{FB}^c$  &    0.0689(35)&   0.0742(7)  & -1.5   \\ &
$s_l^2(Q_{FB})$   &    0.2321(10)  &   0.2314(2)  & 0.7   \\
\hline & $A_{LR}$ & 0.1514(22) & 0.1480(16) & 1.5 \\ SLC & $s_l^2$
($A_{LR}$) & {\it 0.2310(3)}&  {\it  0.2314(2)}  & -1.5
\\
& $A_b$ &  0.9110(250) &  0.9349(1)  & -1.0   \\
& $A_c$ & 0.6300(260)&   0.6683(6)  & -1.5   \\
\hline
LEP II, Tevatron & $m_W$ [GeV] & 80.434(37) & 80.397(23)  & 1.0   \\
& $s_W^2(m_W)$ & {\it 0.2219(7)}  &    &    \\
\hline
Tevatron & $s_W^2$ ($\nu N$) & 0.2255(21)& 0.2231(2) & 1.1 \\
& $m_W(\nu N)$ [GeV] & {\it 80.250(109)} & &  \\
& $m_t$ [GeV]    & 174.3(5.1) &    174.0(4.2) & 0.1\\
\hline
Fit & $m_H$ [GeV]    &   &   $55^{+45}_{-26} $ &  \\
& $\hat{\alpha}_s$ &           & 0.1183(27) &  \\
\hline
$e^+ e^- \to$ hadrons & $\bar\alpha^{-1}$ & 128.878(90) & 128.850(90) & 0.3 \\
\hline
& {\small $\chi^2/n_{d.o.f.}$} & & 21.4/13 & \\
\hline
\end{tabular}

\end{center}
\end{table}

In Table 1 the LEPTOP fit of summer 2001 data is presented. There are
two significant changes in comparison with previous data
presented in Table 2:

1. Due to precision measurement of the cross-section of $e^+ e^-$
annihilation into hadrons in the interval 2-5 GeV at BES the error in
$\bar\alpha \equiv \alpha(M_Z)$ is now two times smaller.
(Following  Electroweak Working Group (EWWG) we use result
\cite{33} though other estimates can be found in the literature as
well);

2. Central value of $M_W$ is now bigger by a half of $\sigma$.

The latter is the main cause for the relaxation of the bound on
heavy extra generations.

Exclusion plot for the number $N_g$ of extra heavy generations is
presented in Fig. 1.

\begin{figure*}[t]
\centering
\includegraphics[width=0.84\textwidth]{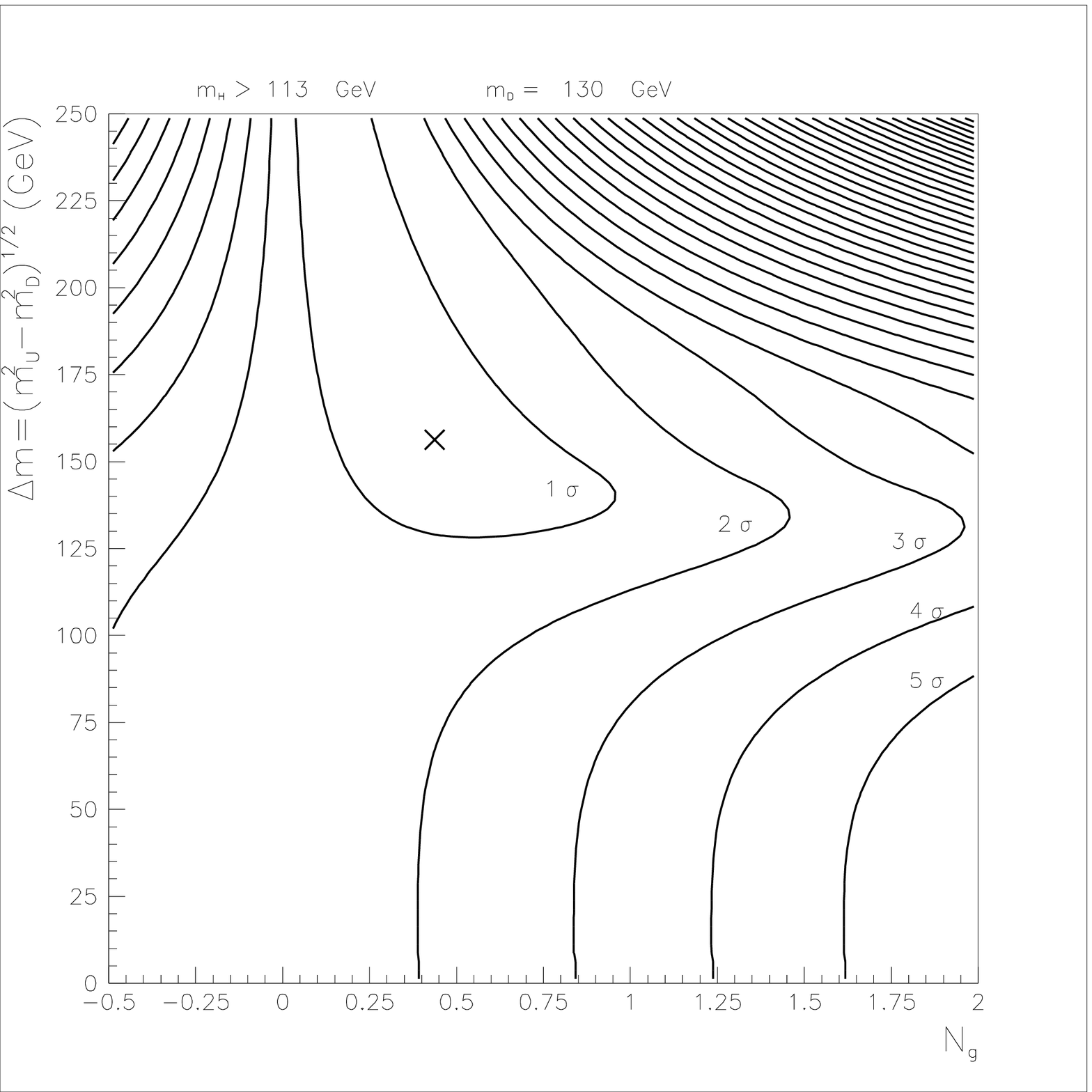}
\caption{\label{FIG1} Exclusion plot for heavy extra generations
with the input: $m_D = m_E = 130$ GeV, $m_U = m_N$. $\chi^2$
minimum shown by cross corresponds to $\chi^2/n_{d.o.f.} =
22.2/12$, $N_g = 0.4$, $\Delta m = 160$ GeV, $m_H = 116$ GeV.
$N_g$ is the number of extra generations. Borders of regions show
domains allowed  at the level $1\sigma, 2\sigma$, etc. }
\end{figure*}

To produce this plot we take $m_D = 130$ GeV -- the Tevatron lower
bound on new quark mass; we use experimental 95\% C.L. bound on higgs
mass $m_H > 113$ GeV \cite{3} and vary $\Delta m = \sqrt{m_U^2 -
m_D^2}$ and number of extra generations $N_g$. (In order to have
two-dimensional plot we arbitrary assumed that $m_N = m_U$ and
$m_E = m_D$; other choices do not change the obtained results
drastically); $\chi^2$ minimum corresponds to unphysical point $N_g
= 0.5$. For $170$ GeV $< m_U < 200$ GeV we get the same quality of
fit in the case $N_g =1$ as that for the SM ($N_g
=0$). In  ref. \cite{55} one can find a statement that extra
heavy generations
are excluded by the precision electroweak data. However, analysis
performed  in \cite{55}
 refers to upper and lower parts of Fig. 1, $\Delta m > 200$ GeV
 and $\Delta m =0$, where the
existence of new heavy generations is really strongly suppressed. This
is not the case for the central part of Fig. 1 ($\Delta m \approx
150$ GeV).

Two heavy generations are excluded at more than $3\sigma$ level.
Nevertheless, two and even three ``partially heavy'' generations
are allowed when
neutral fermions are relatively light, $m_N \simeq 55$ GeV (see
Fig. 2). Using all existing LEP II statistics on the reactions
$e^+ e^- \to \gamma + \nu \bar\nu, \gamma +N \bar N$ in dedicated
search one can exclude 3 ``partially heavy'' generations which
contain such a
light $N$ at a level of $3\sigma$ (see \cite{4}), while one or
even two such generations may exist.

\begin{figure*}[]
\centering
\includegraphics[width=0.84\textwidth]{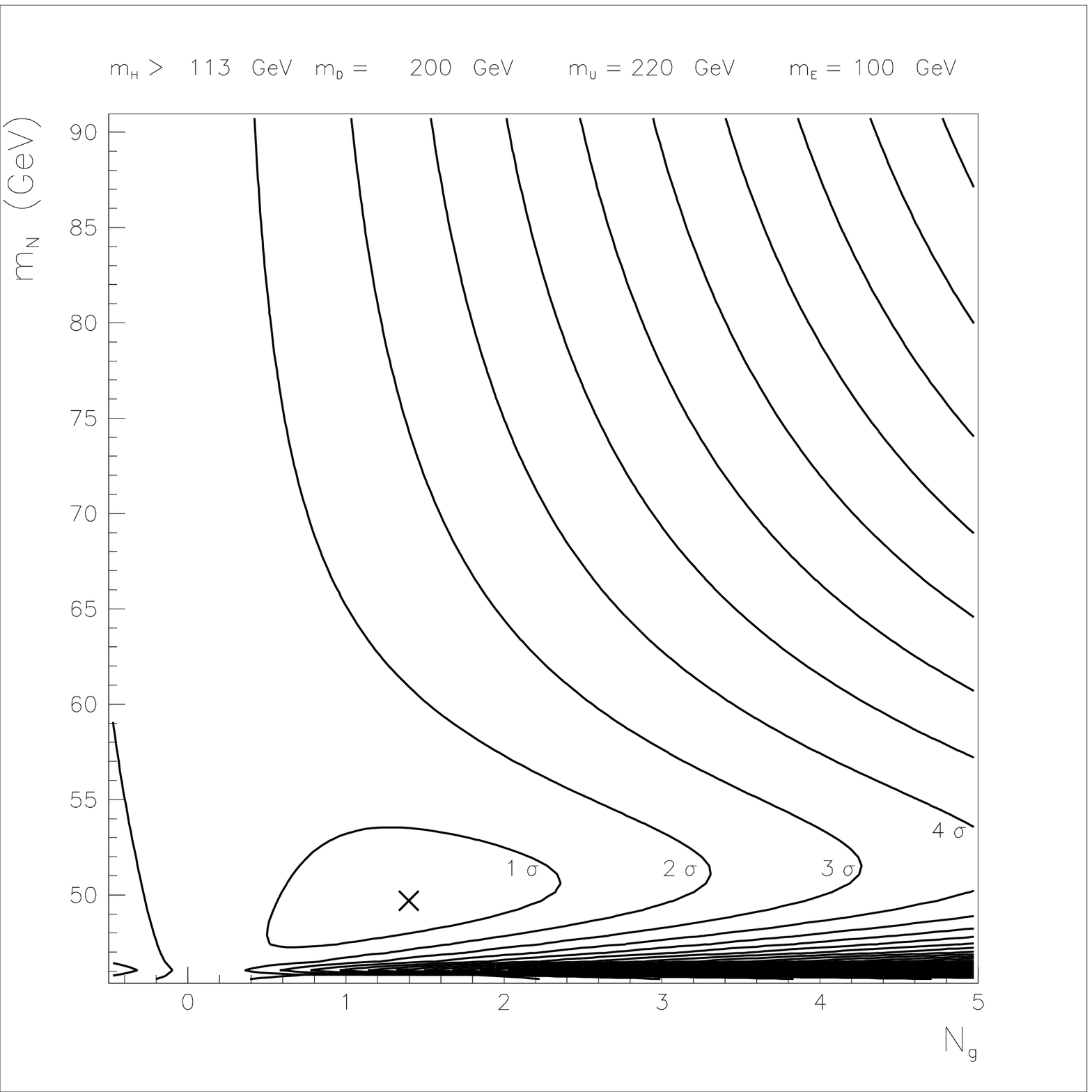}
\caption{\label{FIG2} Exclusion plot for the number of partially
heavy extra generations with light neutral lepton $N$. On
horizontal axis the number of extra generations $N_g$, on vertical
axis -- the mass of the neutral lepton $m_N$. The input: $m_U =
220 $ GeV, $ m_D = 200$ GeV, $m_E = 100 $ GeV. At the minimum
$\chi^2/n_{d.o.f.} = 21.6/12$, $N_g = 1.4$, $m_N = 50$ GeV, $m_H =
116$ GeV. The spectacular behaviour of lines at the bottom of this
figure as well as Fig.4
is caused by the threshold singularity. This singularity must
manifest itself also in the $Z$ lineshape. We have not studied it
because according to experimental data by LEP collaborations on
the emission of initial state bremsstrahlung photon $m_N > 50$ GeV
at 95\% c.l. \cite{4, 444} and the effect at such distance above
threshold is not prominent.
 }
\end{figure*}

The cause of disappearance  of the suppression of extra heavy generations
which existed  in the early data is the contradiction in
description of modern data on $M_W$ and $s_l^2$ in the framework
of SM. The point is that the higgs mass, a free parameter of the SM,
has the following values being
extracted from these observables:
\begin{eqnarray}
(m_W)_{\rm LEP II, Tevatron, NuTeV} & = & 80.428(32) \; {\rm GeV}
\Rightarrow \nonumber \\ & \Rightarrow & m_H = 50^{+50}_{-35} \;
{\rm GeV} \;\; , \label{1}
\end{eqnarray}
\begin{eqnarray}
(s_l^2)_{\rm LEP I, SLAC} & = & 0.23140(15) \;
\Rightarrow \nonumber
\\ & \Rightarrow & m_H = 150^{+75}_{-50} \; {\rm GeV} \;\; .
\label{2}
\end{eqnarray}
$N_g = 0.5$ reduces the contradiction between the two values of $m_H$.
Nevertheless the resulting $\chi^2$ does not improve drastically and
this is due to another ``defect'' of precision data: the
discrepancy between the average value of $s_l^2$ extracted from
pure leptonic measurements and its value from events with hadrons
in final state \cite{3}:
\begin{eqnarray}
 & s_l^2 \nonumber \\
 {\rm Leptons} & 0.23113(21) \nonumber \\
 {\rm Hadrons} & 0.23230(29)
\label{3}
\end{eqnarray}
These 3.3 $\sigma$ difference is the root of poor quality of
the SM fit. The value of hadronic contribution to $s_l^2$ in (\ref{3})
is dominated by very
small uncertainty of  the forward-backward asymmetry
in reaction ~~ $e^+ e^-\to Z\to b\bar b$. According to Table 1
\begin{equation}
(A_{FB}^b)_{\rm exp} = 0.0990(17) \;\; . \label{44}
\end{equation}

 One can question whether such a good accuracy can be obtained in the
analysis of hadronic jets production. Another value of $A_{FB}^b$
can be obtained by multiplying measured at SLAC beauty asymmetry
$A_b$ and leptonic asymmetry $A_l$. Then
\begin{equation}
A_{FB}^b = \frac{3}{4} A_b A_l = 0.1038(25) \;\; . \label{4}
\end{equation}

The number (\ref{4}) differs from (\ref{44}), but nicely  coincides
with the SM fit: 0.1040(8) (see Table 1).

Let us assume following Chanowitz \cite{5} that $A_{FB}^b$
has larger uncertainty than given in Eq. (\ref{44}) and look to what
consequences with respect to extra generations this hypothesis will
lead.\footnote{Another way to resolve situation with $A_{FB}^b$ is
to assume that there exist New Physics contributions to $Zb\bar b$
couplings $g_V^b$ and $g_A^b$. Since in the expression for
$A_{FB}^b$ the corrections to $g_V^b$ and $g_A^b$ are multiplied by
small factor $g_V^e$ they should be large, so they must appear at
the tree level. Also $Z \to b\bar b$ width proportional to
$(g_A^b)^2 + (g_V^b)^2$ should not noticeably change since
$R_b \equiv \Gamma_{Z\to b\bar b}/\Gamma_Z$ is at present
in good agreement with SM fit, see Table 1. In recent paper
\cite{6} inclusion of additional bottom-like heavy quarks with
vector currents is suggested to resolve the discrepancy (\ref{3}).}
If we multiply experimental
uncertainties of $A_{FB}^b$ and $A_{FB}^c$, which are strongly
correlated, by a factor 10, the quality of SM fit improves
drastically: $\chi^2/n_{d.o.f.}$ shifts from 23.8/13 to
10.9/13 and simultaneously one heavy extra
generation becomes excluded at the level of $2.5 \sigma$ (see Fig.
3).

\begin{figure*}[]
\centering
\includegraphics[width=0.84\textwidth]{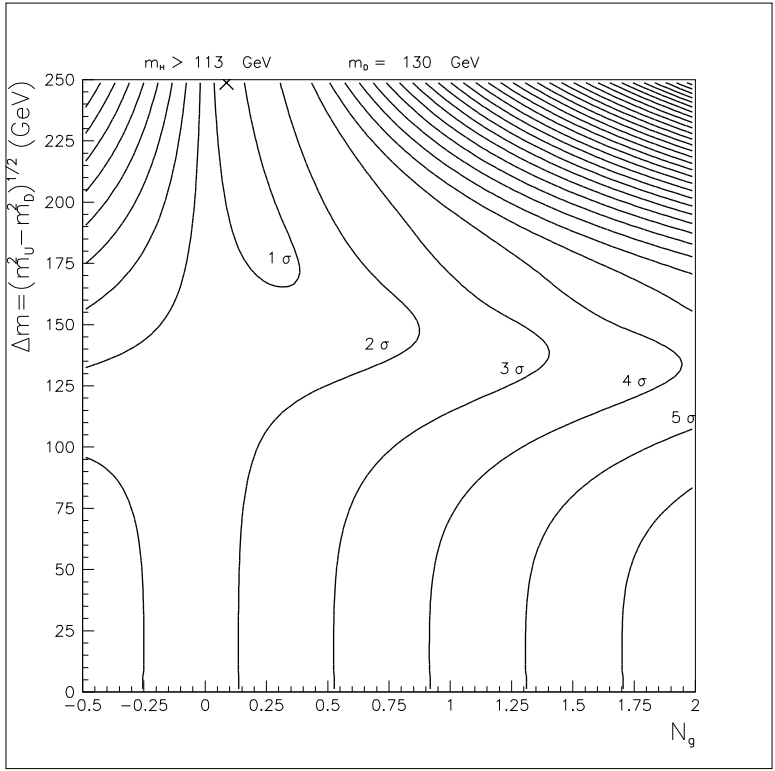}
\caption{\label{FIG3} Exclusion plot for heavy extra generations
with 10 times enlarged errors in  $A_{FB}^b$ and $A_{FB}^c$ with
the input $m_D = m_E = 130$ GeV, $m_U = m_N$. $\chi^2$ minimum is
at the upper border of the Fig., where $\chi^2/n_{d.o.f.} =
11.8/12$, $N_g = 0.1$, $\Delta m = 248$ GeV, $m_H = 116$ GeV. }
\end{figure*}

However, a serious problem arises: it is just $A_{FB}^b$ given by
Eq. (\ref{44}) which pushes $m_H$ to larger values. With our
modification of experimental results on $A_{FB}^b$ and $A_{FB}^c$
the SM fit gives:
\begin{equation}
m_H = 42^{+30}_{-18} \; {\rm GeV} \;\; , \label{6}
\end{equation}
well below modern LEP II bound: $m_H > 113$ GeV, a substantial
trouble for the SM. In case the constraint $m_H > 113$ GeV is
imposed, we get: $m_H = 116^{+15}_{-2}$ GeV, $\chi^2/n_{d.o.f.} =
14.5/14$. What concerns partially heavy extra generations, they
nicely fit the data even with ten times enlarged uncertainties of
$A_{FB}^b$ and $A_{FB}^c$, see Fig. 4.  At both minima in this
Figure $\chi^2/n_{d.o.f.} \simeq 13/12$, while $m_H \simeq 116$
GeV due to the imposed constraint $m_H > 113$ GeV. Without this
constraint $m_H$ drops to $ \sim 40$ GeV, while $\chi^2/n_{d.o.f.} \simeq
10.1/11$ at $N_g = 0.9$, $m_N = 53$ GeV. (The various values of $n_{d.o.f.}$
stems from unconstrained or constrained value of $m_H$ and to
additional parameters $m_N$ and $N_g$ in case of New Physics.)

In the recent paper \cite{99} it was noted that SUSY extension of
Standard Model with light sneutrinos with masses in the range 55-
80 GeV is allowed by precision data and pushes higgs mass to larger
values. ($A_{FB}^b$ was neglected there as well). This might be a
strong indication in favor of light SUSY particles.

 The presence
of new particles is important for production and decay of
higgs.  New heavy quarks  considerably enhance higgs production at
Tevatron and LHC through gluon fusion which should help to
discover this particle \cite{8}. If the decay of higgs into a pair of
neutral leptons is kinematically allowed it will dominate, so that
a  moderately heavy higgs will decay invisibly \cite{10}. At LEP
II the invisibly decaying higgs is excluded almost at the same level as
the SM higgs by missing mass method \cite{11}. Contrary to that
the LHC will look for visible decay modes of higgs. If the
branching ratios of the latter are small the search will be not easy.

\begin{figure*}[]
\centering
\includegraphics[width=0.84\textwidth]{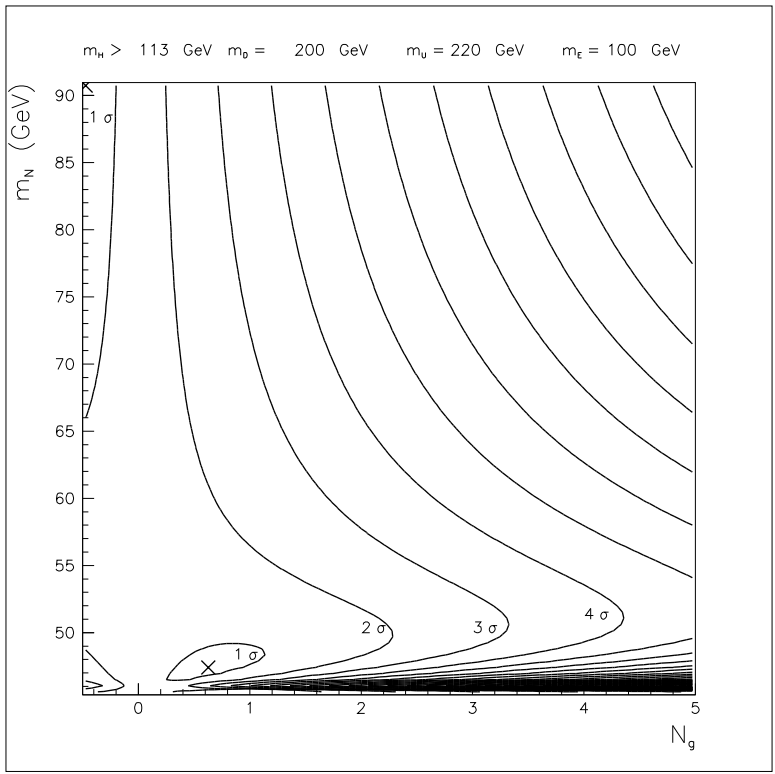}
\caption{\label{FIG4} Exclusion plot for partially heavy extra
generations with 10 times enlarged errors in  $A_{FB}^b$ and
$A_{FB}^c$ with the input $m_D = 200$ GeV, $m_U = 220$ GeV, $m_E =
100$ GeV. Two local $\chi^2$ minima are shown. At the first
minimum $\chi^2/n_{d.o.f.} = 12.4/12$, $N_g = -0.5$, $m_N = 90$
GeV, $m_H = 116$ GeV (see upper left corner of the plot). At the
second minimum $\chi^2/n_{d.o.f.} = 13.1/12$, $N_g =0.6$, $m_N =
48$ GeV, $m_H = 116$ GeV.}
\end{figure*}

We are grateful to A. Olshevsky for providing ref.
\cite{3} and to P.S. Bambade and M. Stanitzki for providing
 ref. \cite{11}.

L.O., A.R. and M.V. are grateful to CERN EP and TH for
hospitality; L.O., V.N. and M.V. were partly supported by RFBR
grant No. 00-15-96562.

\vspace{5mm}

{\bf P.S.} After this paper had been completed a new result for
$s_W^2(\nu N)$ and hence for $m_W(\nu N)$ was published by NuTeV
collaboration \cite{1111}:
$$
s_W^2(\nu N) = 0.2277(17) \;\; ,
$$
$$
m_W(\nu N) = 80.140(80) \;\; .
$$

The new value of $m_W(\nu N)$ differs from $m_W$ measured by LEP
II and Tevatron by 3.7 $\sigma$ and leads to a pull of 2.8 instead
of 1.2 (see Table 1) aggravating the discrepancy. Using the same
procedure as for Table 1 we get:
$$
m_H = 86^{+51}_{-32} \; {\rm GeV} \;\; ,
$$
$$
\chi^2/n_{d.o.f.} = 30.3/13 \;\; .
$$

The influence of the new NuTeV data on the limits on extra
generations, as well as the change of LEPTOP code accounting for
the new NuTeV procedure  of extracting $s_W^2(\nu N)$ will be
discussed elsewhere.

We are grateful to V.Rubakov for providing ref.
\cite{1111}

\vspace{5mm}

{\bf P.P.S.}

As a response to the appearance of this article on hep/ph  H.-J.He
kindly brought to our attention ref. \cite{He}, in which the problem of extra
generations has been considered in a framework of the models with
two and one Higgs doublets. In latter case the results of ref.\cite{He}
could be compared with ours. According to ref.\cite{He}, the $500$ GeV higgs,
 if accompanied by fourth generation, does not contradict the electroweak
precision data. In order to check this statement we made special
LEPTOP runs assuming $m_H=500 $GeV and $N_g=1$. We found that for
certain fixed values of quark and lepton masses the  $\chi^2$ of
the fits with heavy higgs is even better than in the SM. For
example, for
 $m_N = 55 $ GeV, $m_E = 200 $ GeV, $m_U = 130 $ GeV, $ m_D = 130 $ GeV, and
 $m_H = 500 $ GeV  $\chi^2/n_{d.o.f.}= 20.3/14$
which should be compared with $\chi^2/n_{d.o.f.}= 23.8/13$ from Table 1.
Let us note that we do not use S, T, U parametrization of oblique corrections
which is well suited for heavy fermions but not for light ones
(with masses of the order of $M_Z$).

\end{document}